\documentclass[%
 reprint,
superscriptaddress,
 nofootinbib,
 amsmath,amssymb,
 aps,
 prl,
]{revtex4-2}

\usepackage{graphicx}
\usepackage{dcolumn}
\usepackage{bm}
\usepackage{hyperref}

\usepackage[dvipsnames]{xcolor}
\usepackage{orcidlink}
\usepackage{placeins}
\usepackage{color}
\usepackage[dvipsnames]{xcolor}
\usepackage[normalem]{ulem} 

\newcommand{\Aei}{\affiliation{Max Planck Institute for Gravitational Physics (Albert Einstein Institute), D-14467 Potsdam, Germany}}
\newcommand{\Caltech}{\affiliation{Theoretical Astrophysics 350-17, California Institute of Technology, Pasadena, CA 91125, USA}}
\newcommand{\CornellPhysics}{\affiliation{Department of Physics, Cornell University, Ithaca, NY, 14853, USA}}
\newcommand{\Cornell}{\affiliation{Cornell Center for Astrophysics and Planetary Science, Cornell University, Ithaca, New York 14853, USA}}
\newcommand{\CornellLepp}{\affiliation{Laboratory for Elementary Particle Physics, Cornell University, Ithaca, New York 14853, USA}}
\newcommand{\Fullerton}{\affiliation{Nicholas and Lee Begovich Center for Gravitational-Wave Physics and Astronomy, California State University Fullerton, Fullerton, California 92831, USA}}
\newcommand{\Icts}{\affiliation{International Centre for Theoretical Sciences, Tata Institute of Fundamental Research, Bangalore 560089, India}}
\newcommand{\Oberlin}{\affiliation{Department of Physics and Astronomy, Oberlin College, Oberlin, Ohio 44074, USA}}

\begin{document}



\author{Guillermo Lara\,\orcidlink{0000-0001-9461-6292}}
\email{glara@aei.mpg.de} \Aei
\author{Maxence Corman\,\orcidlink{0000-0003-2855-1149}}
\email{maxence.corman@aei.mpg.de} \Aei
\author{Peter James Nee\,\orcidlink{0000-0002-2362-5420}}
\email{peter.nee@aei.mpg.de} \Aei
\author{Harald P. Pfeiffer\,\orcidlink{0000-0001-9288-519X}} \Aei
\author{Nils L.~Vu\,\orcidlink{0000-0002-5767-3949}} \Caltech
\author{Nikolas A.~Wittek\,\orcidlink{0000-0001-8575-5450}} \Aei 

\author{\\Marceline S.~Bonilla\,\orcidlink{0000-0003-4502-528X}} \Fullerton 
\author{Alexander Carpenter\,\orcidlink{0000-0002-9183-8006}} \Fullerton 
\author{Nils Deppe\,\orcidlink{0000-0003-4557-4115}} \CornellLepp \CornellPhysics \Cornell 
\author{Lawrence E.~Kidder\,\orcidlink{0000-0001-5392-7342}} \Cornell 
\author{Prayush Kumar\,\orcidlink{0000-0001-5523-4603}} \Icts 
\author{Geoffrey Lovelace\,\orcidlink{0000-0002-7084-1070}} \Fullerton 
\author{Alexandra Macedo\,\orcidlink{0009-0001-7671-6377}} \Fullerton 
\author{Iago B.~Mendes\,\orcidlink{0009-0007-9845-8448}} \Caltech \Oberlin 
\author{Kyle C.~Nelli\,\orcidlink{0000-0003-2426-8768}} \Caltech 
\author{Mark A.~Scheel\,\orcidlink{0000-0001-6656-9134}} \Caltech 
\author{William Throwe\,\orcidlink{0000-0001-5059-4378}} \Cornell 

\preprint{APS/123-QED}

\title{Signatures from metastable oppositely-charged black hole binaries\\
      in scalar Gauss-Bonnet gravity}

\date{\today}

\begin{abstract}
We conduct numerical simulations of inspiraling, oppositely-charged black holes in the class of scalar-Gauss-Bonnet theories that exhibit spontaneous black hole scalarization.
For quasi-circular, equal-mass binaries near the existence threshold for scalarized solutions, we find a new phenomenon whereby one of the component black holes can suddenly flip the sign of its scalar charge during the inspiral.
We confirm this phenomenon with two independent codes and identify two key signatures thereof: a change in the dominant scalar radiation channel (from dipolar to quadrupolar), and, strikingly,  the introduction of eccentricity in the orbit.
This scenario offers a concrete example of potential nonlinear departures from general relativity in the inspiral of binary black holes in alternative theories of gravity and is of relevance for the development of new tests of gravity.
\end{abstract}

\maketitle



Tests of general relativity (GR) with gravitational waves (GWs) are facing significant challenges.
On the one hand, countless alternative theories of gravity have been proposed (see e.g.~Refs.~\cite{Berti:2015itd, Ezquiaga:2018btd, Shankaranarayanan:2022wbx}), complicating efforts to identify specific deviations from GR.
On the other hand, many of these theories lack concrete predictions for strong and dynamical gravitational fields, further hindering meaningful tests.
Consequently, tests of GR using GWs have focused either on assessing internal consistency with GR, or on searching for parameterized deviations from the GR predictions~\cite{TheLIGOScientific:2016src, Abbott:2018lct, LIGOScientific:2019fpa, LIGOScientific:2021sio}.

If the goal, however, is to directly relate observational departures from GR to new physics, while also accounting for the fact that consistent theories will induce nontrivial correlations among such deviations (see e.g.~Refs.~\cite{TheLIGOScientific:2016src, Perkins:2022fhr, Datta:2022izc, Volkel:2022khh, Maselli:2023khq, Volkel:2020xlc, Gupta:2024gun}), then it is crucial to develop methods for computing predictions in theories \emph{beyond} GR.
Despite substantial progress in leveraging numerical relativity to simulate compact binary coalescences in various theories (see Ref.~\cite{Ripley:2022cdh} for a review), further improvements in accuracy and robustness are needed. These are essential before undertaking large-scale simulation campaigns to build the waveform models required to enable the most stringent tests of gravity~\cite{Yunes:2013dva, Berti:2015itd, Berti:2018cxi, Berti:2018vdi}.
\begin{figure}[]
\includegraphics[width=0.49\textwidth,trim=0 8 0 3.5]{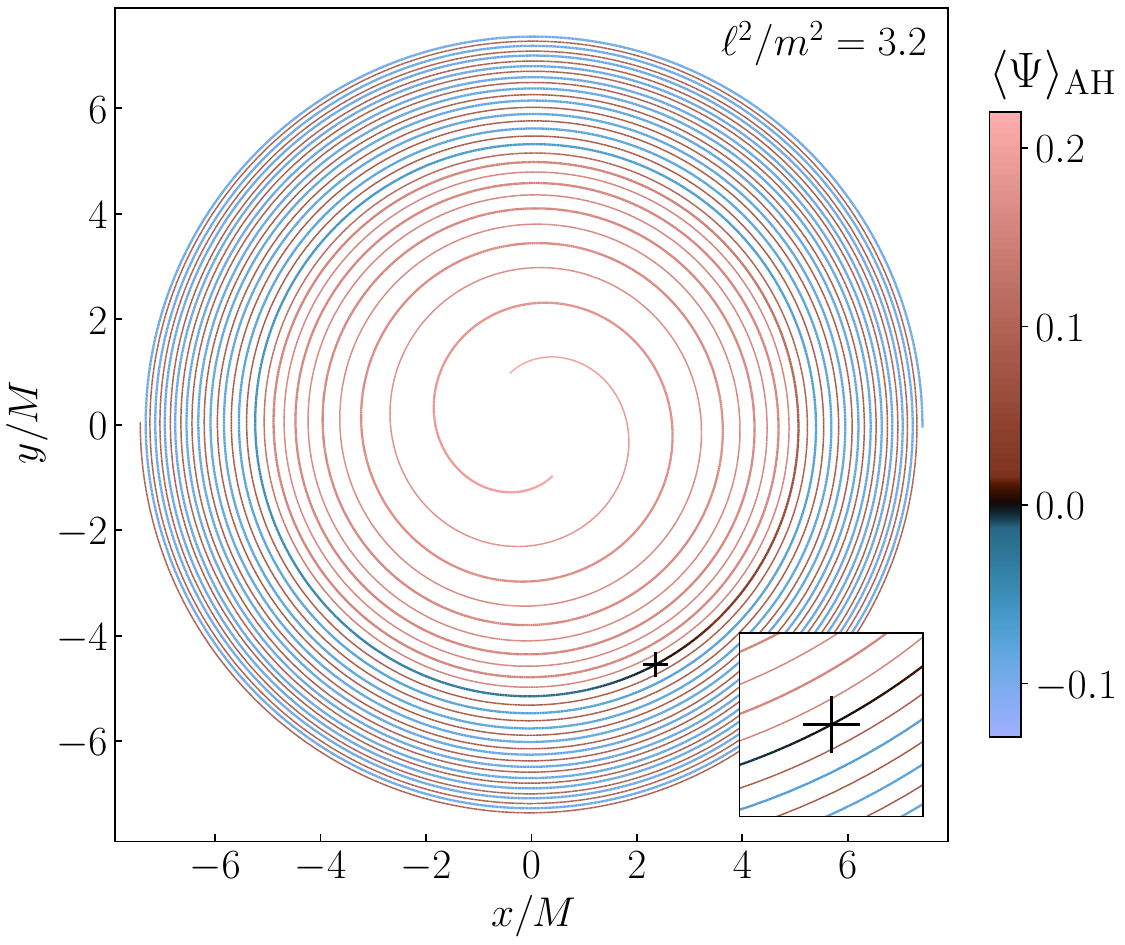}
\caption{\label{fig : flip in trajectory}
        \emph{Charge-flip for an inspiraling quasi-circular equal mass black hole binary in scalar Gauss-Bonnet gravity.}
       The trajectories for black holes A (thick line) and B (thin line) are shown for a simulation with coupling function Eq.~(\ref{eq : coupling gb}) in the test-field limit. The color represents the averaged scalar field at the corresponding apparent horizon. The marker indicates the moment when the charge of black hole A flips sign, with an inset zooming in on this part of the inspiral. Each black hole has mass \(m\), with total mass \(M=2m\).
}
      \end{figure}

In the meantime, numerous beyond-GR waveform examples have already been produced~\cite{Okounkova:2019dfo, Okounkova:2019zjf, Okounkova:2020rqw, East:2020hgw, East:2021bqk, East:2022rqi, Figueras:2021abd, Bezares:2021dma, Corman:2022xqg, AresteSalo:2022hua, AresteSalo:2023mmd,  AresteSalo:2023hcp, Cayuso:2023aht, Held:2023aap, Kuan:2023trn, Corman:2024cdr, Okounkova:2017yby, Witek:2018dmd, Silva:2020omi, Elley:2022ept}.
These provide valuable qualitative guidance to improve and motivate new theory-agnostic tests such as Ref.~\cite{Payne:2024yhk}, where theoretical insight from effective field theory was used to improve inspiral tests of GR.

With this motivation in mind, in this Letter, we report on an unusual example for scalarized black hole (BH) binaries in a scalar-tensor theory of gravity.
Specifically, within the class of scalar Gauss-Bonnet theories that allow for spontaneous scalarization of BHs~\cite{Doneva:2022ewd}, we identify a region in the parameter space where binaries can undergo sudden charge-flips during the inspiral.
In such cases, an initially oppositely-charged binary turns into one where both BHs have charges of the same sign ---see Fig.~\ref{fig : flip in trajectory}.
This charge-flip is accompanied by two key observational signatures:
First, the scalar radiation changes from dipolar to quadrupolar, doubling its frequency.  
Second, the orbit acquires eccentricity, which is in turn imprinted in the  GW radiation, making it potentially observable with GW detectors.

Remarkably, we find that eccentricity can increase in the late-inspiral, challenging the conventional expectation that additional degrees of freedom accelerate circularization via enhanced energy dissipation. 
While this expectation holds in the adiabatic regime~\cite{Cardoso:2020iji}, we show that nonadiabatic effects ---here, a dynamical charge-flip--- can lead to the opposite behavior.
Our results align with other known scenarios where adiabaticity breaks down (e.g.~supernova kicks ~\cite{Colgate1970, Wheeler1975} and resonances or ionization associated with clouds of ultralight scalars~\cite{Boskovic:2024fga, Tomaselli:2024dbw, Tomaselli:2024bdd}), which can also leave nontrivial imprints on the eccentricity of the orbit.
While this phenomenon may help further constrain the theory in a different region of parameter space (extending existing constraints from dipolar emission~\cite{Yagi:2015oca}, Bayesian inference~\cite{Nair:2019iur,Perkins:2021mhb, Lyu:2022gdr, Sanger:2024axs, Julie:2024fwy}, GW propagation speed~\cite{Creminelli:2017sry, Ezquiaga:2017ekz, Baker:2017hug,Sakstein:2017xjx}, and positivity bounds~\cite{Herrero-Valea:2021dry}), its main significance lies in providing a concrete example showcasing how alternative theories of gravity can behave in a clearly distinct way from GR.

\paragraph{\textbf{Theory}.---}

We consider scalar Gauss-Bonnet gravity (a specific example of a four-derivative scalar-tensor effective field theory~\cite{Weinberg:2008hq}), described by the action
\begin{align} \label{eq: action sGB}
      S\equiv \dfrac{1}{16 \pi}\int d^4 x \sqrt{-g}
      \Big[
            \mathcal{R}
            - 2  \nabla_{\!a} \Psi \nabla^{a} \Psi +
            4 \ell^2 f(\Psi) \, \mathcal{G}
            \Big].
\end{align}
Here, \(g = \mathrm{det}(g_{ab})\) is the determinant of the spacetime metric \(g_{ab}\), the scalar field \(\Psi\) is coupled to the Gauss-Bonnet scalar \(\mathcal{G} \equiv \mathcal{R}_{abcd}\mathcal{R}^{abcd} - 4 \mathcal{R}_{ab}\mathcal{R}^{ab} + \mathcal{R}^2\) through a dimensionless function \(f\) and a lengthscale \(\ell\), and we set \(G = c  = 1\).
Varying the action~\eqref{eq: action sGB} with respect to \(\Psi\) and \(g_{ab}\) yields the field equations for this theory:
\begin{align}
      \Box \Psi                                               & = - \ell^2 f'(\Psi) \, \mathcal{G},
      \label{eq: scalar equation}                                                                                                                       \\
      \mathcal{R}_{ab} - \frac{1}{2} \, \mathcal{R} \, g_{ab} & = T_{ab} +  \ell^2 \mathcal{H}_{ab}[ \partial^{\leq 2} g_{cd}, \partial^{\leq 2}\Psi].
      \label{eq: tensor equation}
\end{align}
Here, \(T_{ab} \equiv 2\nabla_{\!a} \Psi\, \nabla_{\!b} \Psi - \nabla_{\!c} \Psi\, \nabla^{c} \Psi \, g_{ab}\), and \(\mathcal{H}_{ab}\) contains up to second-derivative terms of \(g_{ab}\) and \(\Psi\) ---see e.g.~Ref.~\cite{East:2020hgw} for the complete expressions.

Theories with \(f'(\Psi_0) = 0\) and \(f''(\Psi_0)\,  \mathcal{G} < 0\), for \(\Psi_0 \in \mathbb{R},\) admit unique stationary solutions  where \(\Psi=\Psi_0\) and \(g_{ab}\) satisfies the Einstein equations~\cite{Silva:2017uqg}.
However, when \(f''(\Psi_0) \, \mathcal{G} > 0\), such solutions may not be stable and solutions with nontrivial \(\Psi\) can be dynamically preferred.
Indeed, for nonspinning BHs, such \emph{spontaneously scalarized} BHs occur for sufficiently small BH masses, or coupling scale above a certain threshold \(\ell^2/m^2 > \ell^2_\mathrm{th,iso}/ m^2\),  where \(m \) is the mass of the black hole~\cite{Doneva:2017bvd, Silva:2018qhn} ---see also Ref.~\cite{Dima:2020yac} for the case of \emph{spin-induced} scalarization. In the test-field limit \(\ell^2_\mathrm{th,iso}/ m^2 \simeq 2.904\).
In the following, we consider a coupling function that allows for spontaneous BH scalarization:
\begin{align}
      \label{eq : coupling gb}
      f(\Psi) \equiv \dfrac{1}{8}\Psi^2 + \dfrac{\zeta}{16} \Psi^4 ,
\end{align}
where \(\zeta < 0\) is a dimensionless constant, taken here as \(\zeta = - 10\). BH solutions to the theory with Eq.~\eqref{eq : coupling gb} were studied in Refs.~\cite{Silva:2018qhn,Macedo:2019sem,Minamitsuji:2019iwp,East:2021bqk,Silva:2020omi,Doneva:2022byd,Elley:2022ept,Doneva:2017bvd,Silva:2017uqg,Julie:2022huo}.
In addition to their mass \(M\) and spin \(\chi\), BHs acquire an additional \emph{scalar charge} parameter \(q\), encoding information about the asymptotic fall-off of the scalar field, namely \(\Psi(r\to\infty) = \Psi_\infty + q M^2/r + \mathcal{O}(r^{-2})\).

Given our choice of coupling, the action~\eqref{eq: action sGB} remains invariant under the transformation \(\Psi \to -\Psi\). Therefore, a solution to Eqs.~\eqref{eq: scalar equation}-\eqref{eq: tensor equation} with parameters \(\{M, \chi, q\}\) is equivalent to another one with \(\{M, \chi, -q\}\).
For equal-mass binary systems, three configurations are possible: \emph{i)} both charges vanish (\(q_{A} = q_{B}=0\)), \emph{ii)} both charges are non-zero and equal (\(q_A = q_B \neq 0\)), and \emph{iii)} both charges are non-zero, equal in magnitude, but opposite in sign (\(q_A = - q_B \neq  0\)).
In this Letter, we focus on the evolution of binaries of the third kind.

%
\paragraph{\textbf{Methodology}.---}
To study the highly dynamical, nonlinear dynamics of binary systems, it is essential to ensure that solutions to Eqs.~\eqref{eq: scalar equation}-\eqref{eq: tensor equation} are well-behaved and numerically tractable.
As with most alternative theories of gravity, the question of whether scalar Gauss-Bonnet gravity admits a \emph{well-posed} boundary-value and initial-value problem is highly nontrivial~\cite{Papallo:2017qvl, Papallo:2017ddx, Bernard:2019fjb, Julie:2020vov, Witek:2020uzz, Kovacs:2020pns, Kovacs:2020ywu, AresteSalo:2022hua, AresteSalo:2023mmd}, yet crucial for constructing consistent initial data~\cite{Brady:2023dgu, Nee:2024bur} and ensuring stable evolution in time~\cite{Ripley:2019hxt, R:2022hlf, Franchini:2022ukz, Thaalba:2023fmq}.
To address this, we employ two independent numerical relativity codes that implement different methods.

\begin{figure}
  \includegraphics[width=0.47\textwidth,trim=0 10 0 10]{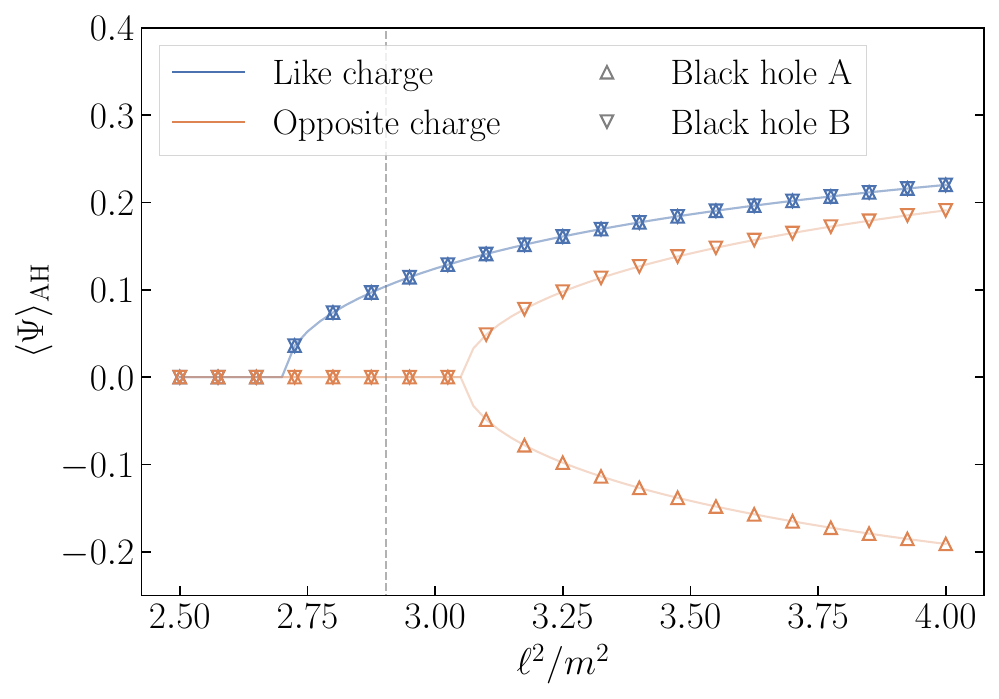}
      \caption{\emph{Initial data calculations of the averaged scalar field at the apparent horizon for quasiscircular equal-mass black hole binaries (test-field).} 
        We consider a quasi-circular binary (non-spinning BHs of equal masses) at fixed orbital frequency \(M\Omega=0.03\). As the strength of the coupling \(\ell^2\) is increased, first a solution with two equal scalar charges appears, followed by a second, independent solution where the BH carry charges of opposite sign. The vertical dashed line indicates the threshold \(\ell^2_\mathrm{th, iso} / m^2 \)~\cite{Silva:2018qhn} for existence of a scalar charge surrounding a single black hole in isolation.
      \label{fig : ID sequence}}
\end{figure}

The first code is \textsc{spectre}~\cite{deppe_2025_14774916}, an open-source numerical relativity code, in which we consider the test-field limit (or decoupling limit) of action~\eqref{eq: action sGB}, where the backreaction of the scalar field onto the metric is neglected ---i.e.~Eq.~\eqref{eq: tensor equation} reduces to \(\mathcal{R}_{ab} = 0\).
We construct initial data for quasi-circular BH binaries in equilibrium by solving the augmented extended conformal thin-sandwich equations~\cite{Nee:2024bur} using the elliptic solver~\cite{Vu:2021coj, Vu:2024cgf} within \textsc{spectre}.
We scan the parameter space by producing sequences of initial data sets for different orbital parameters (separation \(D\), orbital frequency \(\Omega\) and radial velocity) and coupling values \(\ell^2\).
This approach implicitly defines an \emph{adiabatic} approximation, assuming instantaneous equilibrium and neglecting effects on timescales shorter than the orbital dynamics.
Compared to parameter exploration based solely on evolution codes, this method offers significant computational efficiency, enabling us to identify regions of parameter space relevant for charge-flips (see Fig.~\ref{fig : ID sequence}).

Interesting initial data sets are evolved in the test-field limit within \textsc{spectre} using the system described in Ref.~\cite{Lara:2024rwa}. The metric is evolved using a generalized harmonic formulation~\cite{Lindblom:2005qh} in damped harmonic gauge~\cite{Varma:2018evz}. Further eccentricity reduction is performed following Refs.~\cite{Pfeiffer:2007yz, Buchman:2012dw}.
The initial data and evolution codes in \textsc{spectre} employ a discontinuous-Galerkin scheme ---see Refs.~\cite{Teukolsky:2015ega, Vu:2024cgf, Lovelace:2024wra} for further details.

A second code is used to investigate the effect of backreaction and its impact on the GW signal. It implements~\cite{East:2020hgw} the full equations~\eqref{eq: scalar equation}-\eqref{eq: tensor equation}, rewritten in the modified generalized harmonic formulation~\cite{Kovacs:2020pns, Kovacs:2020ywu}. We make similar choices for the gauge and numerical parameters as in Ref.~\cite{East:2020hgw}, except that we find the scalarized BHs evolved here also benefit from the addition of long-wavelength constraint damping, obtained by setting \(\rho\! =\! -0.5\) in Eq.~(2) of~\cite{East:2020hgw}.
For initial data, we start from quasi-circular binary vacuum GR solutions constructed using the \textsc{TwoPunctures} code~\cite{Ansorg:2004ds,Paschalidis:2013oya,code_repo_tp}. We then add a small Gaussian scalar perturbation centered on the BHs, following the procedure
in Ref.~\cite{East:2021bqk}. In all cases presented here, we use an initial amplitude of \(\Psi_0 = 0.008\).
We have verified that smaller amplitudes yield the same results, and that the error induced by not solving the constraint equations including the perturbation is negligible.
See the End Matter for details on resolution and convergence.


\paragraph{\textbf{Initial data sequences}.---}

In isolation, a BH of mass \(m\) will spontaneously scalarize and acquire a scalar charge $q$ when the coupling \(\ell^2/m^2\) exceeds the threshold value \(\ell^2_\mathrm{th, iso}/m^2\).
The presence of another BH with scalar charge of the same sign enhances the scalar field around each BH.
In a binary BH, this mutual amplification lowers the existence threshold for scalarization, effectively expanding the region of parameter space where BHs can support scalar hair.
In particular, for an equal mass binary with \(m_A=m_B=m\) and same-sign scalar charges, \(\ell_{\rm th}^2/m^2<\ell^2_{\rm th,iso}/m^2\).
For evolving binaries, this effect leads to \emph{dynamical} scalarization~\cite{Palenzuela:2013hsa, Taniguchi:2014fqa} of BHs~\cite{Julie:2023ncq}, where initially non-scalarized BHs acquire scalar charges of the same sign as they get closer later in the inspiral.

Here, we study the converse scenario. When the companion BH has a charge of opposite sign, we find that the scalar field is instead \emph{weakened} around each BH, \emph{increasing} the 
scalarization threshold. For equal-mass binaries, this means $\ell^{2}_\mathrm{th}/m^2 > \ell^{2}_\mathrm{th, iso}/m^{2}$.
Figure~\ref{fig : ID sequence} illustrates this effect in the test-field limit by showing the averaged scalar field on the horizon \(\langle \Psi \rangle_\mathrm{AH}\) (a proxy of the scalar charge), as a function of \(\ell^2/m^2\) for sequences of quasi-equilibrium binary BH initial data at fixed orbital frequency \(M \Omega = 0.03\), equivalent to a fixed separation.
For this orbital frequency, the scalarization threshold shifts by approximately \(5\%\) in either direction: lower for like-charged BHs (blue curve), and higher for oppositely-charged BHs (red curve).
We also find (not shown) that in both cases, the magnitude of the threshold shift increases as the separation \(D\) decreases.
Thus, at fixed \(\ell^2/m^2\), BHs can cross the scalarization threshold 
during inspiral as their separation decreases.
For binaries with opposite scalar charges, our results suggest a phenomenon contrary to dynamical scalarization: there exists a region of parameter space in which such binaries will 
lose their scalar hair or become unstable during inspiral.

\begin{figure}
      \includegraphics[width=0.47\textwidth]{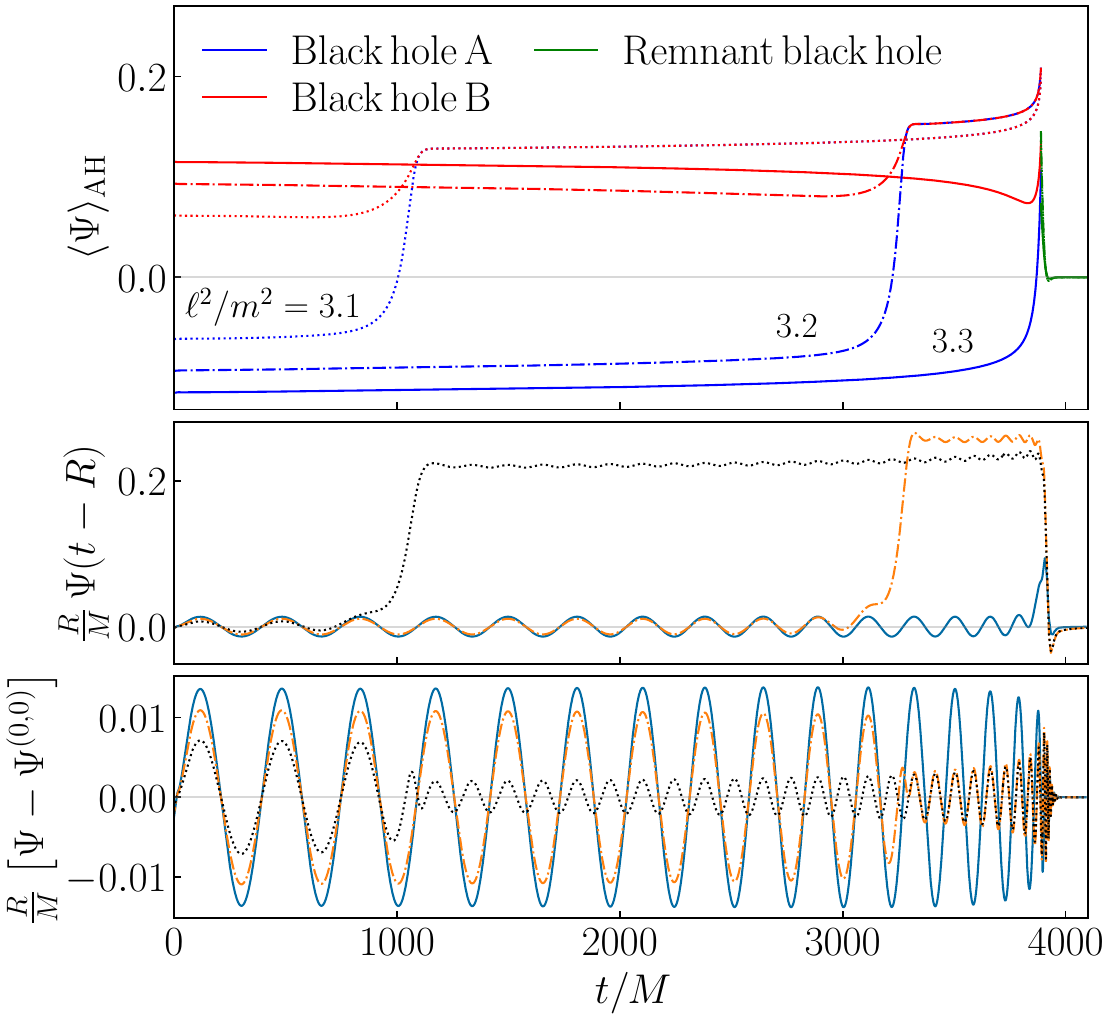}
      \caption{\emph{Scalar radiation and scalar field at the apparent horizon (test-field).} \textbf{Top:} Average scalar field at the horizon of both black holes for different values of the coupling \(\ell^2/m^2\), where \(m\) is the Christodoulou mass of the component black holes.
      In the test field case, we empirically find charge-flips in the parameter region \(\ell^2_\mathrm{th, iso} / m^2 < \ell^2/m^2 \lesssim 3.3\), where \(\ell^2_\mathrm{th, iso} / m^2  \).
      \textbf{Middle:} Finite radius \(R/M=170\) (where \(M = 2m\)) scalar  wave at a point in the equatorial plane ---we only consider modes \(\ell \leq 2\). \textbf{Bottom:} Same as the middle panel but substracting the monopole contribution to highlight the doubling of the frequency after charge-flip.
      }
      \label{fig : scalar waveform dec cases}
\end{figure}

\paragraph{\textbf{Evolutions}.---}

Guided by the insights provided by our initial data exploration, we perform full nonlinear numerical evolutions targeting the region of parameter space where oppositely-charged binaries become ``metastable''.
We consider equal-mass systems with coupling values just above the shifted existence threshold at the initial separation.
As these systems evolve, they initially maintain stable scalar charges of opposite sign. 
However, as the separation decreases and the binaries cross the existence threshold for oppositely-charged configurations, one of the two BHs flips its scalar charge and the system migrates to the like-charged branch.
Due to the symmetry of the system under \(\Psi \to - \Psi\), there is no preferred sign for the like-charged configuration. In practice, the truncation error in the initial data determines the sign (see End Matter for further details), which is why the initial data prediction of complete scalar hair loss is not realized.

\begin{figure}
    \includegraphics[width=0.47\textwidth]{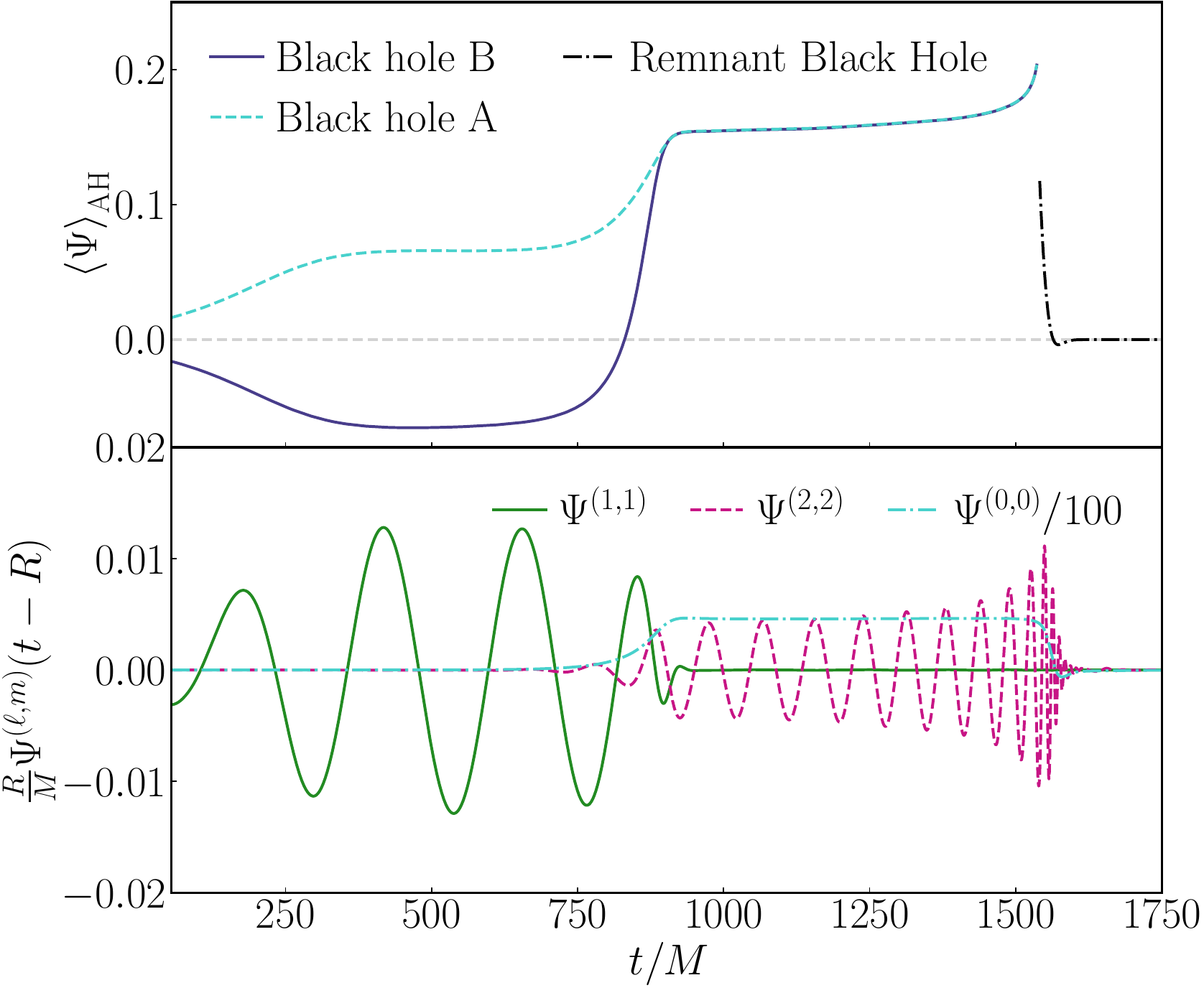}
      \caption{\emph{Scalar radiation and scalar field at the apparent horizon (with backreaction).}
        \textbf{Top:} Average scalar field at the horizon of both black holes for a value of the coupling \(\ell^2/m^2 = 3.14\) and initial separation \(D/M = 11.6\), where \(m\) is the Christodoulou mass of the component black holes and \(M=2m\).
      \textbf{Bottom:}  Scalar radiation modes for some spherical harmonic modes \((\ell,m)\) extracted at radius \(R/M=100\). When BH B flips to a positive scalar charge, the scalar radiation becomes quadrupolar.
      } 
      \label{fig : scalar waveform coupled}
\end{figure}

The charge-flip effect is shown in the top panels of Fig.~\ref{fig : scalar waveform dec cases} (test-field limit) and Fig.~\ref{fig : scalar waveform coupled} (fully-coupled system).
The computationally less expensive simulations in the test-field limit allow us to explore different values of the coupling constant, see Fig.~\ref{fig : scalar waveform dec cases}:
As \(\ell^2/m^2\) approaches the existence threshold for isolated BHs, \(\ell^2_\mathrm{th, iso}/m^2 \)~\cite{Silva:2018qhn}, the flip of BH A occurs earlier in the inspiral ---formally, up to spatial infinity exactly at that threshold.
Conversely, as \(\ell^2/m^2\) increases, the flip occurs progressively closer to merger.  Because of the test-field approximation, all simulations in Fig.~\ref{fig : scalar waveform dec cases} merge at the identical time \(t_{\rm merger}\sim 3900 \,M\); 
after merger, the remnant BH of mass \(M_f\) \emph{descalarizes}, as it is below the scalarization threshold, i.e.~\(\ell^2/M^2_f < \ell^2_\mathrm{th, iso}/M^2_f \)~\cite{Silva:2020omi}. 

Since the scalar-dipole energy flux is proportional to \(\left(q_{A}-q_{B}\right)^2\) [see e.g.~Eq.(73) of Ref.~\cite{Yagi:2015oca}], the dominant scalar radiation channel during the earlier part of the inspiral is shut down during the transition to the like-charged configuration.
The scalar radiation is subsequently dominated by quadrupolar emission at twice the orbital frequency, as seen in the lower panel(s) of Fig.~\ref{fig : scalar waveform dec cases} (plotting the scalar field for an observer in the orbital plane) and Fig.~\ref{fig : scalar waveform coupled} (plotting a mode decomposition).
A monopolar component is also visible since \(q_A + q_B > 0\) after the charge-flip and
until the remnant descalarizes.

\begin{figure}
  \includegraphics[width=0.47\textwidth]{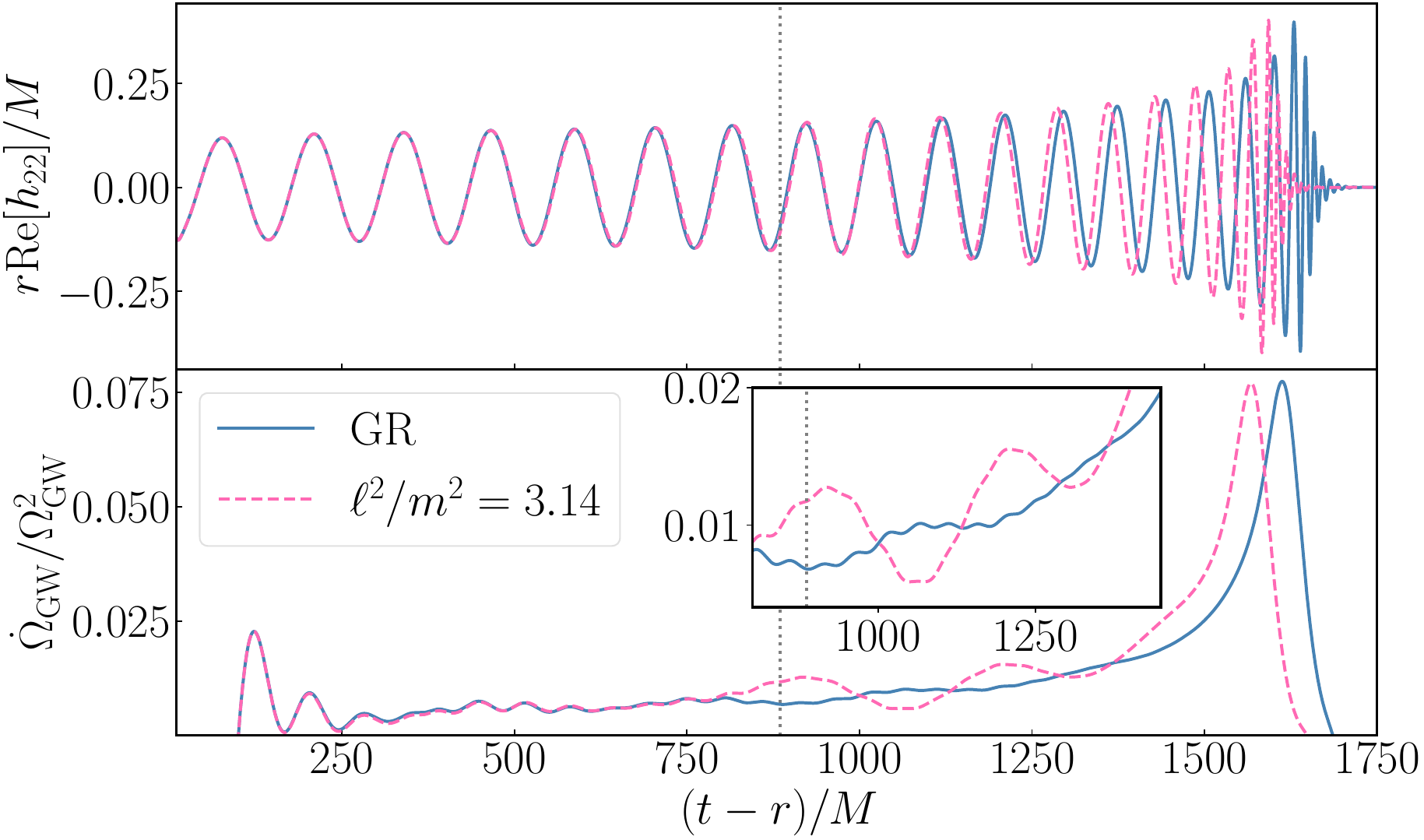}
      \caption{\emph{Gravitational radiation and time derivative of
        gravitational wave frequency.}  \textbf{Top:} In pink (solid line), finite radius
        \(R/M=100\) gravitational radiation modes for \(\ell^2/m^2 =
        3.14\) and initial separation \(D/M = 11.6\) when compared to
        a quasi-circular system in GR with the same orbital parameters
        (blue and dashed).  \textbf{Bottom:} Time derivative of GW frequency normalized by
        square of GW frequency. The grey dotted line corresponds to
        time at which the scalar field on horizon of black hole A in
        Fig.~\ref{fig : scalar waveform coupled} goes through zero.
        The charge-flip introduces eccentricity in the orbit and
        causes the merger to occur earlier.  }
      \label{fig : grav waveform}
\end{figure}

The evolution including scalar backreaction (results of which are shown in Fig.~\ref{fig : scalar waveform coupled}) enables a comparison with standard GR.
As expected, the scalarized binary spirals in more quickly, owing to the additional energy dissipation in the scalar radiation channel.  Moreover, the charge-flip leads to the introduction of \emph{eccentricity} (roughly of \(e \sim 10^{-2}\)) in a system that was nearly quasicircular, with \(e \simeq 1.6 \times 10^{-3}\) at the start of the simulation.
Figure~\ref{fig : grav waveform} illustrates both effects. In particular, the bottom panel shows the eccentricity-driven oscillations in the GW frequency (in pink), which set in during the charge-flip, and from which \(e\) can be estimated.
In blue, we compare to a quasicircular system in GR. The corresponding gravitational waveforms are shown in the top panel of Fig.~\ref{fig : grav waveform}.
The increase in eccentricity may be associated with the decrease in the Christodoulou mass of each component BH during the charge-flip, which is on the order of \(\lvert \Delta m \rvert /m \sim 0.7\, \% \).
(As in previous studies [e.g.~Refs.~\cite{Ripley:2019irj, Cayuso:2020lca, East:2021bqk}], the Null Convergence Condition does not hold here, allowing the horizon area of each individual BH to \emph{decrease}.)
This behavior is reminiscent of orbital eccentricity introduced in binary stars when one component suddenly sheds mass, as in a supernova.

\paragraph{\textbf{Conclusions}.---}

In this Letter, we uncover a particularly interesting and unexpected phenomenon in scalar Gauss-Bonnet gravity: a sign-flip in the scalar charges of equal-mass, oppositely charged binary BHs.
This effect can be viewed as the inverse counterpart of the well-known dynamical scalarization process~\cite{Palenzuela:2013hsa, Taniguchi:2014fqa, Khalil:2022sii, Julie:2023ncq}.
Using our newly developed initial data code~\cite{Nee:2024bur}, we chart the region of parameter space where oppositely-charged binaries cease to admit equilibrium solutions during the inspiral.
We then perform full nonlinear evolutions of these systems using two independent numerical relativity evolution codes: \textsc{spectre}~\cite{Lara:2024rwa, deppe_2025_14774916} and that of Ref.~\cite{East:2020hgw}.
Our results show that, instead of completely shedding their scalar hair, such binaries can undergo a sudden charge-flip in one of the component BHs at any point during the inspiral.  
The charge-flip changes the dominant scalar radiation channel from dipolar to quadrupolar, and introduces \emph{eccentricity} in the GW signal. 
The modification of the GW emission thus opens up the possibility of detecting imprints of 
scalarized BHs, even if the scalar radiation itself is not directly observable.
 
Our results also bring closure to earlier work in the literature.
In Ref.~\cite{Julie:2022huo}, the adiabatic inspiral of oppositely-charged binaries was studied in the Post-Newtonian framework. However, in that work, the evolution of the binary became inaccessible below a certain critical separation.
Here, we have shown that the final fate of such systems is to become like-charged binaries.
Other examples in the literature where non-gravitational radiation channels can change during the inspiral include binaries interacting through  massive dark photons~\cite{Alexander:2018qzg, Owen:2025odr}.
However, in that case, the change in vector radiation is not associated to a change in nature of the inspiralling compact objects, but is due to the Yukawa interaction mediated by the massive vector.
More broadly, we expect this behaviour to have analogues in other alternative theories of gravity where compact objects are not described by unique solutions. The most obvious candidate are oppositely-charged neutron star (NS) binaries in the Damour-Esposito-Far\`ese theory~\cite{Damour:1996ke}, the theory in which dynamical scalarization was originally discovered~\cite{Palenzuela:2013hsa, Taniguchi:2014fqa}.
In fact, Ref.~\cite{Sennett:2017lcx} argued using an effective point-particle model that oppositely charged neutron star binaries are disfavored with respect to the like-charged binaries at close separations. To our knowledge, however, no numerical simulations have confirmed this yet.

Arguably, binaries experiencing charge-flips in scalar Gauss-Bonnet gravity are expected to occur only within a narrow region of the parameter space;
e.g.~assuming beyond-GR effects enter at a scale \(\ell \sim M_\star\), the relevant
mass range around \(M_\star\) is roughly \(\lvert \Delta m \rvert / M_\star \sim \lvert \Delta (\ell^2) \rvert / 2 \ell^2_\mathrm{th, iso} \sim 10 \%\), where \(\Delta (\ell^2)/M_\star^2 \sim 0.4\) is the spread in coupling values observed in Fig.~\ref{fig : scalar waveform dec cases}.
Whether such binaries can form through astrophysically viable channels remains an open question.

A central goal of this work, however, is to inform the GW community developing tests of GR about unexpected phenomena in alternative theories of gravity. In this regard, our work highlights two key points.
First, strategies must be developed to identify such state transitions, so that agnostic tests of the inspiral phase can recognize them.
In the spirit of inspiral-merger-ringdown consistency 
tests~\cite{Sampson:2014qqa, TheLIGOScientific:2016src, Abbott:2018lct, LIGOScientific:2019fpa, LIGOScientific:2021sio}, one possibility could be to compare early and late inspiral phases, provided sufficient signal-to-noise ratio is retained in both.
Second, it is vital to extend current tests of GR to account for eccentric effects,
as assumptions about the circularization of LIGO-Virgo-KAGRA (\(1-100\, M_\odot\)) binaries may not generically hold when gravity is not described by vacuum GR.


\paragraph{\textbf{Acknowledgments}.---}
The authors would like to thank Felix-Louis Juli\'e, Raj Patil, Hector O.~Silva, Thomas Sotiriou and Jan Steinhoff for fruitful discussions.
Computations were performed on the Urania HPC systems at the Max Planck Computing and Data Facility.
This material is based upon work supported by the National Science Foundation under Grants No.~PHY-2309211; No.~PHY-2309231; and No.~OAC-2209656 at Caltech, and No.~PHY-2407742; No.~PHY-2207342; and No.~OAC-2209655 at Cornell. Any opinions, findings, and conclusions or recommendations expressed in this material are those of the author(s) and do not necessarily reflect the views of the National Science Foundation. This work was supported by the Sherman Fairchild Foundation at Caltech and Cornell. This work was supported in part by NSF awards PHY-2208014 and AST-2219109, the Dan Black Family Trust, and Nicholas and Lee Begovich at Cal State Fullerton. P.K.~acknowledges support of the Department of Atomic Energy, Government of India, under project no.~RTI4001, and by the Ashok and Gita Vaish Early Career Faculty Fellowship at the International Centre for Theoretical Sciences.

\bibliography{paper}

\FloatBarrier
\subsection*{End Matter}
%


\paragraph{\textbf{Appendix A: Convergence in test-field limit}.---}

\begin{figure}[]
      \includegraphics[width=0.47\textwidth]{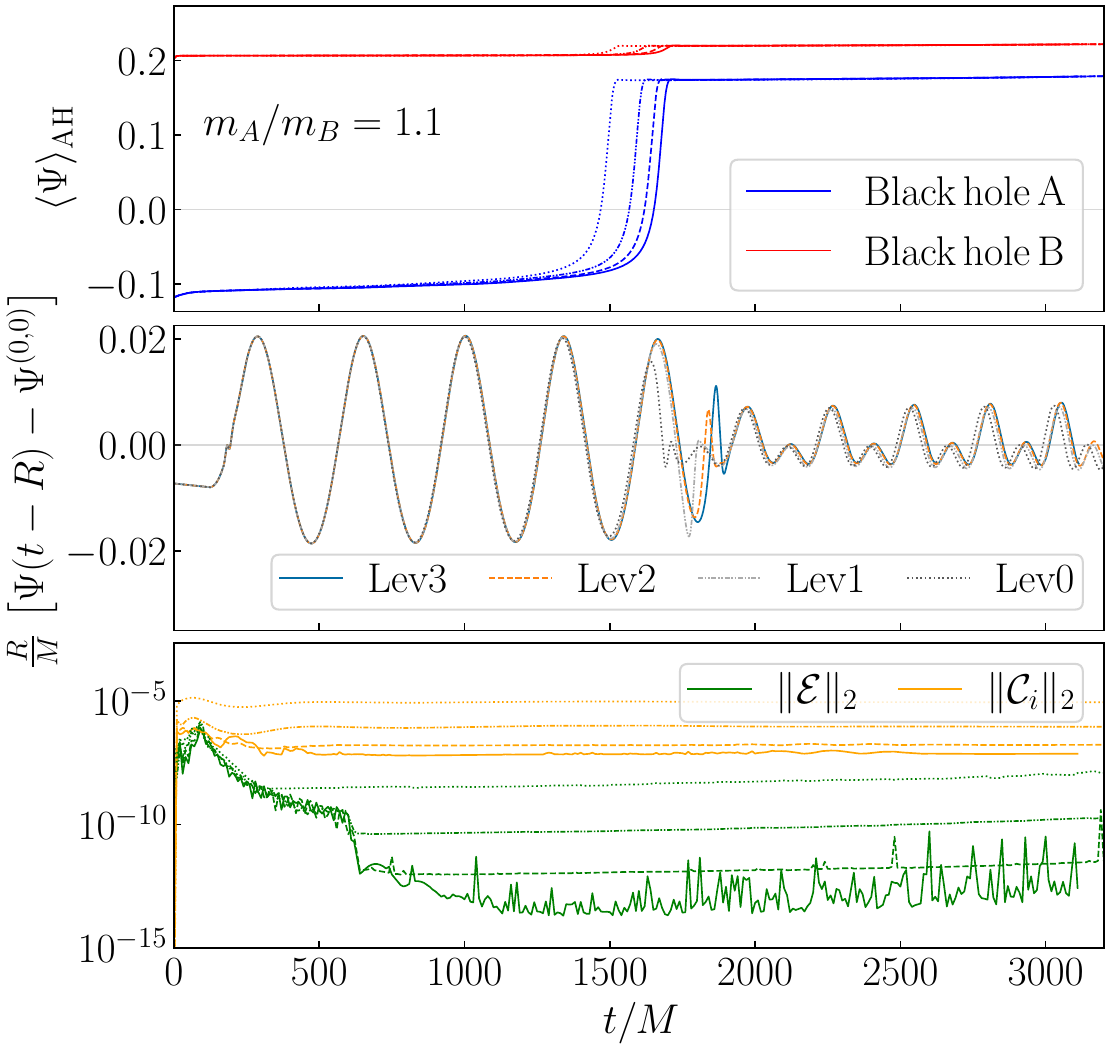}
      \caption{
      \emph{Convergence test for an unequal-mass, charge-flipping binary with mass-ratio $m_A/m_B=1.1$.}
        \textbf{Top:} Scalar field  \(\langle \Psi \rangle_{\mathrm{AH}}\), averaged over each of the BH horizons. \textbf{Middle:} Value of the scalar field (with monopole removed) at a point in the equatorial plane. \textbf{Bottom:} violation of the constraint energy \(\mathcal{E}\) (see Eq.~(19) of Ref.~\cite{Lovelace:2024wra}) and first-order constraint of the scalar system \(C_{i} \equiv \partial_{i} \Psi - \Phi_{i}\).
        Plotted are four different resolutions Lev0 (lowest) to Lev3 (highest).
        The data in the middle panel is computed using the spherical harmonic modes with \(\ell=1,2\), and the lower panel shows \(L_2\) norms over the entire computational volume.}
      \label{fig : conv test unequal mass ratio}
      \end{figure}

For simplicity, the main text focuses on equal-mass binaries, for which the truncation error in the initial data selects one of the two equally likely possibilities for the endstate after the charge-flip: either \(q_{A} = q_{B} > 0\) or \(q_{A} = q_{B} < 0\).
The inherent randomness in the symmetry breaking makes convergence tests difficult for the system discussed in the main text.  Figure~\ref{fig : conv test unequal mass ratio} presents a charge-flipping binary at mass-ratio $m_A/m_B=1.1$.  In this system,   the \emph{physical} asymmetry in the binary causes the more massive BH to experience the charge-flip, and we obtain numerically convergent solutions.
The data in Fig.~\ref{fig : conv test unequal mass ratio} is obtained in the test-field limit with \textsc{spectre}.
%


\paragraph{\textbf{Appendix B: Convergence of the fully coupled system}.---}

\begin{figure}
      {\includegraphics[width=0.49\textwidth]{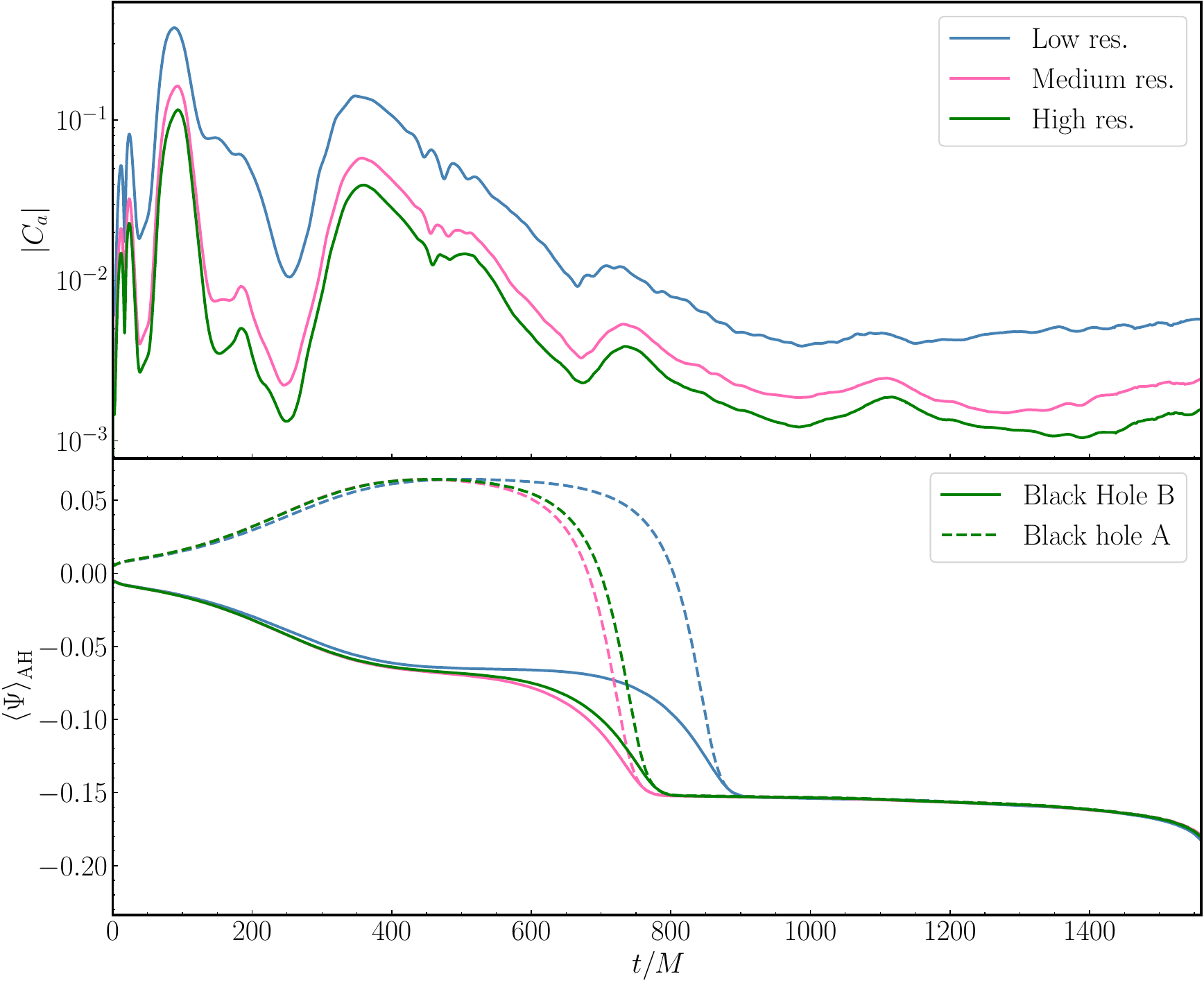}}
          \caption{\emph{Convergence test for the fully coupled equal mass system.} 
          \textbf{Top:} Integrated norm of the constraint violation $|C_a|$ 
          as a function of coordinate time (in units of the total mass) for the 
          equal-mass binary black hole merger with \(\ell^2/m^2 = 3.14\) and initial 
          separation \(D/M = 11.6\) presented in Fig.~\ref{fig : grav waveform},
          at three resolutions. 
          The medium and high resolutions have 1.5 and $2 \times$ the resolution
          of the low resolution simulation. We find third order convergence as expected.
          \textbf{Bottom:} Average scalar field at the horizon of both black holes 
          at the corresponding three resolutions.}
          \label{fig:conv_norm_equal_mass}
          \end{figure}

Figure~\ref{fig:conv_norm_equal_mass} presents a convergence test for the simulation shown in Fig.~\ref{fig : grav waveform}.
The top panel shows the integrated modified generalized harmonic constraint violation \(|C_a|\) as a function of coordinate time for three resolutions. This quantity shows third-order convergence to zero, as expected.
The lowest resolution has eight levels of refinement. The finest level has a linear grid spacing of \(dx \sim 0.016 M\) and each successive level has a linear grid spacing that is twice as coarse. 
The medium and high resolutions are 1.5 and \(2 \times\) as high, respectively. The lowest resolution is the default resolution for the results presented in this work.
The sign of the charges after the charge-flip (here, \(q_A = q_B < 0\)) can be inferred from the scalar field averaged over each BH horizon ---bottom panel of Fig.~\ref{fig:conv_norm_equal_mass}.

\begin{figure}
      \includegraphics[width=0.49\textwidth]{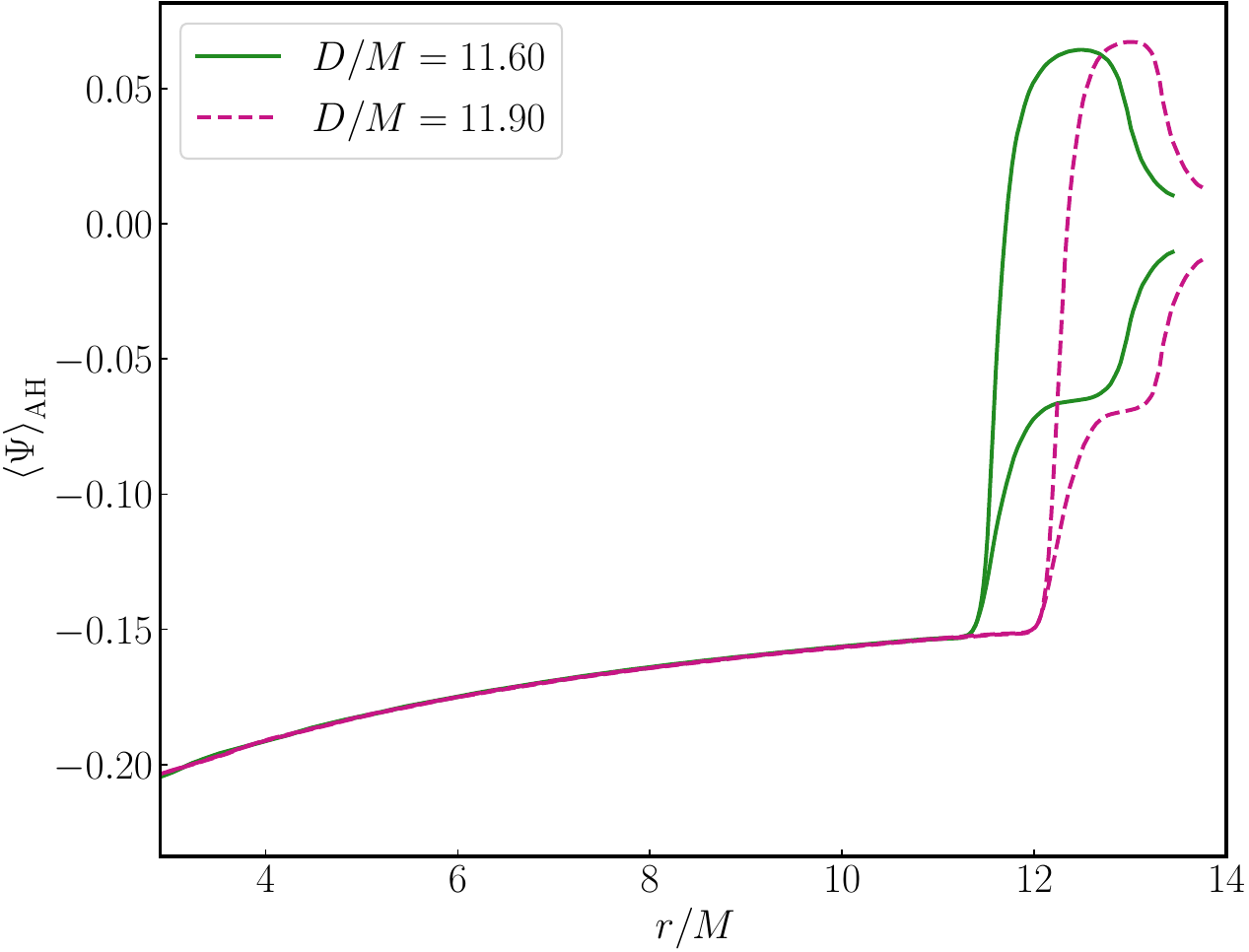}
      \caption{
        \emph{Fully coupled equal mass system starting at different initial separation.} 
          Average scalar field at the horizon of both black holes 
          for a value of the coupling \(\ell^2/m^2 = 3.14\) but two initial separations, \(D/M = 11.6\) and \(11.9\) as a function of coordinate separation.}
          \label{fig:equal_mass_larger_sep}.
          \end{figure}

Finally, Fig.~\ref{fig:equal_mass_larger_sep} shows the dependence of our results with respect to the initial separation. 
For \(\ell^2/m^2 = 3.14\), we evolve equal-mass, oppositely-charged systems starting from initial separations \(D/M = 11.6\) and \(11.9\).
The orbital separation at which the flip occurs changes slightly with the initial separation.
This initial data dependence is not unexpected since (as mentioned in the main text), for equal-mass binaries, the truncation error in the initial data breaks the symmetry of the system and selects the exact charge-flip time and sign of the like-charged system after the transition.

      %
%

\end{document}